\documentclass[aps,pre,preprint,superscriptaddress,groupedaddress,ams]{revtex4-1}
\usepackage{graphicx}  
\usepackage{bm}        
\usepackage{amssymb}   
\usepackage{physics}
\usepackage{tikz}
\usepackage{gensymb}
\usetikzlibrary{shapes,snakes}
\usetikzlibrary{shapes.geometric}
\usepackage{pgfplots}
\usepackage{chngcntr}
\usepackage{dsfont}
\usepackage[caption = false]{subfig}
\usepackage{hyperref}
\usepackage{xcolor}
\usepackage{float}
\usepackage[utf8]{inputenc}
\usepackage[english]{babel}
 \usepackage{amsthm}
 \usepackage{relsize}

\theoremstyle{definition}

\newcommand{\newc}{\newcommand}
\newc{\beq}{\begin{equation}}
\newc{\eeq}{\end{equation}}
\newc{\kt}{\rangle}
\newc{\br}{\langle}
\newc{\beqa}{\begin{eqnarray}}
\newc{\eeqa}{\end{eqnarray}}
\newc{\pr}{\prime}
\newc{\longra}{\longrightarrow}
\newc{\ot}{\otimes}
\newc{\rarrow}{\rightarrow}
\newc{\h}{\hat}
\newc{\bom}{\boldmath}
\newc{\btd}{\bigtriangledown}
\newc{\al}{\alpha_k}
\newc{\be}{\beta_k}
\newc{\ld}{\lambda}
\newc{\sg}{\sigma}
\newc{\p}{\psi}
\newc{\eps}{\epsilon}
\newc{\om}{\omega}
\newc{\mb}{\mbox}
\newc{\tm}{\times}
\newc{\ra}{\rightarrow}
\newc{\non}{\nonumber}
\newc{\ul}{\underline}
\newc{\hs}{\hspace}
\newc{\longla}{\longleftarrow}
\newc{\ts}{\textstyle}
\newc{\f}{\frac}
\newc{\df}{\dfrac}
\newc{\ovl}{\overline}
\newc{\bc}{\begin{center}}
\newc{\ec}{\end{center}}
\newc{\dg}{\dagger}
\newc{\T}{\mathcal{U}}
\newc{\Tp}{\mathcal{V}}
\newc{\J}{\mathsf{J}}
\newc{\sfL}{\mathsf{L}}
\newc{\C}{\mathsf{C}}
\newc{\B}{\mathsf{M}}
\newc{\V}{\mathsf{V}}

\begin{document}

\widetext

\title{Out-of-time-ordered correlators and quantum walks}

\author{Sivaprasad Omanakuttan}
\email[]{somanakuttan@unm.edu}
\affiliation{Department of Physics and Astronomy, University of New Mexico, MSC07-4220, Albuquerque, New Mexico 87131-0001, USA}
\author{Arul Lakshminarayan}
\email[]{arul@physics.iitm.ac.in}
\affiliation{Department of Physics, Indian Institute of Technology Madras, Chennai 600036, India}

\date{\today}

\begin{abstract}
Out-of time-ordered correlators (OTOC) have recently attracted significant attention from the physics of many-body systems, to 
quantum black-holes, with an exponential growth of the OTOC indicating quantum chaos. Here we consider OTOC in the context of coined discrete quantum walks, a very well studied model of quantization of classical random walks with applications to quantum algorithms. Three separate cases of operators, variously localized in 
the coin and walker spaces, are discussed in this context and it is found that the approximated behavior of the OTOCs is well described by simple algebraic functions  in all these three cases with different time scale of growth. The quadratic increase of OTOC signals the absence of quantum chaos in these simplest forms of quantum walks. 
 \end{abstract}

\maketitle

\section{Introduction}

Out-of-time-ordered correlators (OTOC) were first introduced by Larkin and Ovchinnikov in the theory of superconductivity\cite{larkin}; however,  it has come into prominence recently in very different contexts, finding applications in many different fields including quantum chaos, high-energy, and condensed matter physics \cite{maldacena2016bound,moudgalya2018operator,shenker2014black,
polchinski2016spectrum,roberts2015localized,
shenker2015stringy,swingle2016measuring,
hashimoto2017out,PhysRevB.97.161114,PhysRevA.95.011601,kitaev2018soft,lakshminarayan2018out,PhysRevE.98.052205}. It is simplest to understand it as the growth of non-commutativity of operators and hence the classical counterparts are related to Poisson brackets which can 
grow with time exponentially for chaotic systems. Hence it has led to discussions of the ``butterfly-effect" in unconventional settings such as black holes and spin models 
\cite{maldacena2016bound,aleiner2016microscopic,shenker2014black,roberts2014two,
shenker2015stringy}

Let the two unitary operators $A$ and $B$ be such that they commute at time $t=0$. Consider the quantity $F(t)$ directly related to the Euclidean or Frobenius norm of the commutator between the evolved operator $A(t)$ and the operator $B$, given by   
\begin{equation}
\begin{aligned}
\label{eq:Expression for OTOC 1}
F(t)=&\frac{1}{2D}\|\left[A(t),B\right]\|_F^2\\=&\frac{1}{2D}\text{Tr}\left([A(t),B]^{\dagger}[A(t),B]\right)\\=& 1-\frac{1}{D}\text{Re}\left[ \text{Tr}(A(t)^{\dagger}B^{\dagger}A(t)B)\right],
\end{aligned}
\end{equation}
where $D$ is  the dimension of the operators under consideration, $B(t=0)=B$ and $A(t)=e^{iHt}A e^{-iHt}$ is the evolution of the operator in the Heisenberg picture (with $\hbar=1$).  The second term in the above expression is the out-of-time-ordered correlator (other definitions need not include a full trace and may be an expectation value with respect to states such as the thermal state \cite{shenker2015stringy}). This governs the time evolution of the commutator, therefore define $1-F(t)$ as 
\beq
\label{eq:OTOC}
C_4(t)=\frac{1}{D} \text{Re}\left[ \text{Tr}(A(t)^{\dagger}B^{\dagger}A(t)B)\right].
\eeq
If, as we assumed, the operators $A$ and $B$ commute at $t=0$, $F(t)$ increases from $0$ in a manner that  is a signature of the dynamics and can eventually saturate after a long time. 

Quantum walks \cite{aharonov1993quantum,kempe2003quantum,
childs2002example} come primarily in two flavors, namely the discrete and continuous time versions, and has found potential applications in quantum computation \cite{childs2009universal,lovett2010universal}, quantum search algorithms \cite{childs2004spatial,aaronson2003quantum} and various other fields \cite{chandrashekar2008optimizing,zhang2016creating,
di2014quantum,genske2013electric,muraleedharan2018quantum}. The discrete quantum walk is the object of study in this paper. Due to its connections to classical random walks it is of interest to 
see how ``chaotic" it may be with the measure of the OTOC. Independently it is an interesting dynamical question to consider.

Discrete quantum walk consists of a walker and a coin, and its Hilbert space is a tensor product of their individual spaces. As in the classical case, the walker moves in the lattice based on the coin states. The position translation operator $\T$  acting on the position eigenkets $| n\kt$ shifts them:
\begin{equation}
\label{eq:position space1}
\T\vert{n}\rangle=\vert{n+1\pmod N}\rangle,
\end{equation} 
where $0\leq n<N$ and $N$ is the total number of lattice sites and we assume periodic boundary condition. Defining $\omega=\exp\left(\frac{2 \pi i }{N}\right)$, the momentum states $\vert{\tilde{k}}\rangle$ are eigenvectors of $\T$:
\begin{equation}
\label{eq:position space2}
  \T\vert{\tilde{k}}\rangle=\omega^{-k}\vert{\tilde{k}}\rangle,
 \end{equation} 
while the  momentum translation operator $\Tp$ satisfies the following relation with $l=k+1\,\text{mod} N $:
\begin{equation}
\label{eq:momentum space1}
\Tp |{\tilde{k}}\kt=|\tilde{l}\kt,\;\; \Tp|{n}\kt=\omega^{n}|{n}\kt.
\end{equation}
The commutation relation between the position translation operator and the momentum translation operator is given by the standard Weyl relation \cite{schwinger1960unitary}:
\begin{equation}
\label{eq:commutation relation}
\Tp \T=\omega \, \T \Tp.
 \end{equation}
The unitary operator corresponding to the discrete quantum walk is given as \cite{nayak2000quantum}\cite{kempe2003quantum}
\begin{equation}
U(\theta)=\left(|{0}\kt\br{0}|\otimes \T +|{1}\kt\br{1}|\otimes \T^{\dagger}\right)(\overset{\sim}{U_{\theta}}\otimes \mathtt{I}_N).
\end{equation}
Here $N$ is the size of the lattice, $0 \leq \theta\leq\pi/2$, and 
\begin{equation}
\overset{\sim}{{U_{\theta}}}=\begin{pmatrix}
\cos \theta  & \sin \theta\\
\sin \theta & -\cos \theta
\end{pmatrix},
\end{equation}
is the coin dynamics that occurs in a two dimensional Hilbert space
spanned by the orthogonal states $\{ |0\kt, |1\kt\}$.
Due to the translational symmetry of the walk, $U(\theta)$ is block diagonal in the momentum basis of the walker $|\tilde{k}\kt$ \cite{nayak2000quantum} and is given as $U(\theta)=\bigoplus_{k} U_k(\theta)$ where,
\begin{equation}
\label{eq:block diagonal representation of U_csw}
U_k(\theta)=\left( \begin{array}{cc}
\cos \theta\, \omega^k & \sin \theta\, \omega^{k} \\
\sin \theta\, \omega^{-k}&-\cos \theta\, \omega^{-k}
\end{array} \right),
\end{equation}
and $0 \leq k < N-1$. When $\theta=\pi/4$, this corresponds to 
the well-studied Hadamard coin \cite{kempe2003quantum}.

There is a growing interest to understand the behavior of OTOCs where there is no quantum chaos, the recent study of OTOCs in the integrable quantum Ising model being an example. Power law growth of OTOCs are observed in integrable system as opposed to the expected exponential growth observed in nonintegrable systems exhibiting quantum chaos\cite{lin2018out,kukuljan2017weak,riddell2018out,huang2017out,sun2018out}. The coined discrete quantum walk is a simple model that has been posed also a a coherent version of classical random walks. 

When $\theta=\pi/4$ in the coin dynamics, measurement of the coin at each step leads to the classical unbiased random walk \cite{kempe2003quantum}. Thus it could be of interest to investigate how quantum chaotic this walk is. The OTOC is a measure of how two perturbations applied at different times do not commute and is a measure of the sensitivity
of the system. The lack of quantum chaos in this and a phase space version of it was discussed recently in \cite{omanakuttan2018quantum} and motivates one to investigate the behavior of OTOCs in various versions of quantum walks, we start therefore with the standard and simplest model in this paper, and indeed exhibit power-laws in the OTOCs, studying them both analytically and numerically. 

In  the discrete coined quantum walk, as mentioned above a coin and a walker space exist. In this paper a separate analysis of the OTOCs when both the operators ($A$ and $B$) are in the coin space is carried out in the section II, similar analysis for the walker space is in section III,  while in section IV we consider the case in which one operator is in the coin space and the other is in the walker space. We will conclude the paper by a summary and possible future directions in section V.

\section{ OTOC\smaller{s} for the Coin-Coin case}

This section is devoted to the study of OTOC for the case in which both the operators, $A$ and $B$ as discussed in the introduction, belong to the coin space. The model of discrete quantum walk discussed in this paper has a coin with a two-dimensional Hilbert space; however there are generalizatios with higher dimensional coin spaces \cite{lakshminarayan2003random,panahiyan2018one,lorz2018photonic}
which are of independent interest.

Consider  $ A= B=H \otimes \mathds{1}_N$, which are in the coin space, where $H$ is the Hadamard gate or operator 
\begin{equation}
\label{eq:Hadamard Matrix}
H:=\frac{1}{\sqrt{2}}\begin{pmatrix}
1 & 1\\
1 & -1
\end{pmatrix}.
\end{equation}
Since $H \otimes \mathsf{1}_{N}$ is Hermitian as well as unitary the expression of the $C_4(t)$ in the Eq.~\eqref{eq:OTOC} becomes, 
\begin{equation}
\label{eq:OTOC for the coin-coin case}
\begin{aligned}
C_4(t)=&\frac{1}{2N} \text{Re}\left[ \text{Tr}\left(\left[A(t)A\right]^2\right)\right].
\end{aligned}
\end{equation}
However, the time evolution of the operator $A$ in the Heisenberg picture is given in terms of $U_k(\theta)$ as,
\begin{equation}
\label{eq:Time evolution of operator in the Coin-Coin Case }
A(t)=\sum_{k=0}^{N-1}U_k^{-t} (\theta)H U_k^t(\theta),
\end{equation}
which gives, 
\begin{equation}
\label{eq:OTOC in the Coin-Coin Case using block diagonal matrix}
F_{CC}(t)=1-\frac{1}{2N}\sum_{k=0}^{N-1}\text{Tr}[\left(U_k^{-t}(\theta)H U_k^{t}(\theta) H\right)^2].
\end{equation}
The marginal cases of $\theta=0$ and $\theta=\pi/2$ are naturally special but give rise to some interesting non-trivial behaviors for $F_{CC}$.
 \subsection{Marginal Cases}
 When the coin is $\sigma_z$ ($\theta=0$) and $\sigma_x$ ($\theta=\pi/2$) further simplification of the Eq.~\eqref{eq:OTOC in the Coin-Coin Case using block diagonal matrix} in the coin-coin case using block diagonal matrix is possible and the simplified expression of $F_{CC}(t)$ can be found, see App.~(\ref{app:coin space}) for details of the calculation. For the $\sigma_z$ coin, 
\begin{equation}
\label{eq:Coin-coin case for theta=0}
F_{CC}(t)=\frac{5}{4}-\frac{1}{4}\left[4(-1)^t \, \delta(2 t\,\text{mod}N,0)+\delta(4 t\,\text{mod}N,0)\right].
\end{equation}
Thus $F_{CC}(0)=0$, while $F_{CC}(t\geq 1)=5/4$, till one of the Kronecker deltas click again which 
is at time scales of the order of $N$ and is hence very long. 
If the initial state is $\ket{\psi}_c\ket{\phi}_w$, then the state 
the state after $n$ time steps is given as,
 \begin{equation}
\ket{\psi(t=n)}=\bra{0}\ket{\psi}_c\ket{0}\T^n \ket{\phi}_w+(-1)^n\bra{1}\ket{\psi}_c\ket{1}\left(\T^{\dagger}\right)^n\ket{\phi}_w.
\end{equation}
Hence the initial coin state's  contribution to the overall state remains same after time $t=1$, which is in-fact reflected in the behavior of $F_{CC}$.\\
When the coin dynamics is given by $\sigma_x$,
\begin{equation}
F_{CC}(t)=\begin{cases}
0\,\; \text{if $t$ is even, }\\
5/4\; \text{for $t$ odd.}
\end{cases}
\label{eq:Coin-coin case for theta=pi/2}
\end{equation}
Considering the same initial state as before 
 \begin{equation}
 \ket{\psi(t=1)}=\bra{1}\ket{\psi}_c\ket{0}\T \ket{\phi}_w+\bra{0}\ket{\psi}_c\ket{1}\T^{\dagger}\ket{\phi}_w,
 \end{equation}
 and
  \begin{equation}
 \ket{\psi(t=2)}=\ket{\psi(t=0)}.
 \end{equation}
 Hence there are only two possible configurations for the system like a classical coin (no superposition of the configuration is allowed). This back and forth nature similar to the two level atom as discussed in App.~(\ref{app:two level atom}) results in the rapid oscillation of the $F_{CC}$.  

\subsection{Non-Marginal Case}
For the non-marginal cases, an alternate expression for  Eq.~\eqref{eq:OTOC in the Coin-Coin Case using block diagonal matrix} is obtained using the unitary matrix ${\hat{U}~=\bigotimes_{k=0}^{N-1}T_k}$, where $T_k $ diagonlizes $U_k$, and is given as,
\begin{equation}
\label{eq:OTOC for the coin-coin case after diagonalization of U_k 2}
C_4(t)=\frac{1}{2N}\sum_{k=0}^{N-1}\Tr[\left(D_k^{-t}(\theta)A_k D_k^{t}(\theta)A_k\right)^2],
\end{equation}
where $A_k=T_kHT_k^{\dagger}$ and, 
\begin{equation}
\label{eq:Diagonalization of block matrix}
D_k=T_k^{\dagger}U_kT_k=\begin{pmatrix}
\exp (i\alpha_k)   &0\\
0 & -\exp (-i\alpha_k) 
\end{pmatrix},
\end{equation}
with $\alpha_k=\sin^{-1}\left(\cos\theta \sin 2\pi k/N\right)$. Let  
\begin{equation}
\label{eq:transformation of Hadamard matrix}
A_k:=\begin{pmatrix}
a_{11}(k) &a_{12}(k)\\
a_{21}(k)&a_{22}(k)
\end{pmatrix},
\end{equation}
however, $A_k$ is Hermitian as well as unitary and is similar to $H$ which yields $\Tr (A_k)=0$ and $\det(A_k)=-1$, hence  $a_{11}(k)=-a_{22}(k)$, $a_{12}(k)=a_{21}(k)^{*}$, and $|a_{12}(k)|^2+a_{11}^2(k)=1$. Defining 
\begin{equation}
R_k(t):=\Tr[\left(D_k^{-t}(\theta)A_k D_k^{t}(\theta)A_k\right)^2],
\end{equation}
 $\left(C_4(t)=\sum_k R_k(t)\right)$  can be represented in terms of $a_{11}(k)$ and $a_{12}(k)$ as,
\begin{equation}
\label{eq:OTOC for the coin coin case -k th term}
\begin{aligned}
R_k(t)=&2\left[a_{11}^4(k)+(-1)^ta_{11}^2(k)|a_{12}(k)|^2(-2+4\cos 2\alpha_k t)\right]\\ &+2 |a_{12}(k)|^4\cos 4\alpha_k t.
\end{aligned}
\end{equation}
Hence the expression of $C_4(t)$ becomes,
\begin{widetext}
\begin{equation}
\label{eq:OTOC for the coin-coin after hadamard approximation}
\begin{aligned}
\C_4(t)=2\sum_{k=0}^{N-1}\left[a_{11}^4(k)+(-1)^t(-2+4\cos 2\alpha_k t) a_{11}^2(k)|a_{12}(k)|^2 +|a_{12}(k)|^4\cos 4\alpha_k t\right].
\end{aligned}
\end{equation}
\end{widetext}
 An exact evaluation  of $a_{11}(k)$ and $a_{12}(k)$ is possible for the case of $\theta=\pi/4$ and is given in the App.~(\ref{app:Explicit Calculation of a11}). However the resultant sums appear formidable, and so we resort to using their averages over $k$ (given ahead in Eq.~\eqref{eq:Average values of a11}) further followed by conversion of the summation to an integral which yields the approximation
\begin{equation}
\label{eq:OTOC for the coin coin case- integral approximation 1}
\begin{aligned}
C_4(t)&\approx 8-\frac{11}{\sqrt{2}}+\frac{(1-\frac{1}{\sqrt{2}})}{N}\int_{k=0}^{N}dk \cos 4 \alpha_k t\\
&+(-1)^t\frac{(-8+6\sqrt{2}) }{N}\int_{k=0}^{N}dk\cos 2 \alpha_k t .
\end{aligned}
\end{equation}
Since $\sin^{-1} x \approx x$ for small values of $x$, this yields $ \alpha_k \approx\cos \theta \sin 2 \pi k/N=1/\sqrt{2} \sin 2 \pi k/N$, which in turn results in the final approximation of the OTOC as,
\begin{equation}
\label{eq:bessel approximation for coin}
\begin{aligned}
F_{CC}(t)\approx &\frac{11-7\sqrt{2}}{\sqrt{2}}+\left(\frac{1-\sqrt{2}}{\sqrt{2}}\right)J_0(2\sqrt{2}\, t)\\&+(-1)^{t+1}(-8+6\sqrt{2})J_0(\sqrt{2} \,t).
\end{aligned}
\end{equation}
%

\begin{figure}
   \subfloat{\includegraphics[width = 3.2in]{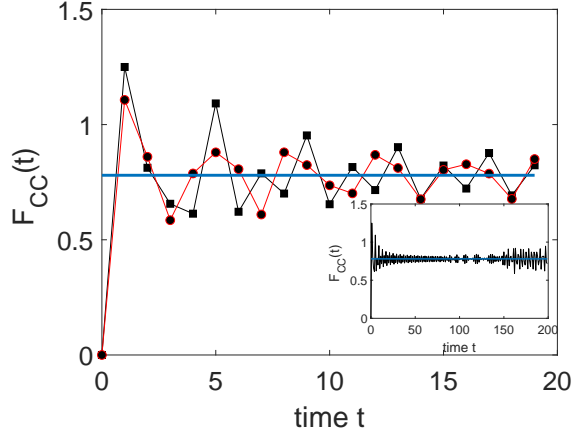}} 
   \caption{Growth of OTOC with time for the coin-coin case is given for  $\theta=\pi/4$, the line with square symbol is the $F_{CC}$ and the line with circle (red color) is the one obtained using the approximations, the inset gives the long time behavior of $F_{CC}(t)$ for  a lattice of size $N=200$. The blue line is $y=0.78$, which shows the algebraic damping. }
 \label{fig:OTOC for the coin Case for CSW}
 \end{figure}

The growth of $F_{CC}(t)$ with $t$ is given in Fig.~(\ref{fig:OTOC for the coin Case for CSW}), $F_{CC}(t)$ shows rapid oscillatory nature for this case in the initial time steps which tends to saturating behavior at later time. Since $J_0(\beta\, t)=J_0(-\beta\, t)$,  $F_{CC}(t)$  has time reversal symmetry.   
This equation shows that one relevant time scale for the coin-coin OTOC is of $O(1)$ at which it starts to oscillate. However these oscillations show
an algebraic damping from the Bessel decay as $1/\sqrt{t}$ to the constant value $(11-7 \sqrt{2})/\sqrt{2}\approx 0.78$, for large $N$. The OTOC in a two level system with time independent Hamiltonian is derived in the App.~(\ref{app:two level atom}), which shows similar behavior as the coin-coin case discussed in this section; however, it is periodic. The $F_{CC}$ in the quantum walk can be considered as a reflection of this fact that the two-level coin leads to the oscillations but due to interaction with the walker these damp out and eventually $F_{CC}$ saturates.

 Since the walk analogous to the classical case happens in the lattice (walker space) and the distinction from the classical case has been the major breakthrough of the discrete quantum walk, in the next section as stated in the introduction, operators in the walker space is  considered and the OTOCs is studied in detail and it is found that the OTOCs gave insights into the lack of quantum chaos in the quantum walk and also about localization of information in the walker space.
  
\label{section:OTOC in coin space}
\section{ OTOC\smaller{s} for the Walker-Walker  case}
\label{section:OTOC in walker space}
When both the operators are nontrivial in the walker space,
a natural choice would be when they are the momentum translation operator $\Tp$, as the complementary position translation operator 
$\T$ commutes with the quantum walk dynamics. Therefore,
if we consider the operators, $A=B=\mathds{1}_2\otimes \Tp$, the growth of  OTOCs, $F_{WW}(t)$,in this case is an interesting quantity. The time evolution of $A$  in the momentum basis of the walker $(|\tilde{k}\kt)$  is given as,
\begin{equation}
\label{eq:Evolution of the operator in the walker-walker space}
\br{\alpha \tilde{k}}|A(t)|{\beta \tilde{k}'}\kt=\br{\alpha }|U_{k}^{-t}(\theta)U_{k'}^t(\theta)|{\beta} \kt\delta_{k,k'+1}^{N}.
\end{equation}
Where $|\alpha\kt, \,|\beta \kt $ are arbitrary coin states and  $k,k^{'}=0,1,2,....,N-1$ gives the lattice (walker) indices. The operation of modulo on the momentum indices is represented by $\delta^{N}$. Using this, a compact formula for the OTOC is found as a sum over different momentum sectors:
\begin{equation}
\label{eq:OTOC for the walker-walker space }
F_{WW}(t)=1-\frac{1}{2N}\sum_{k=0}^{N-1}\text{Tr}[U_{k-2}^{-t}(\theta)U_{k-1}^t (\theta)U_{k}^{-t}(\theta)U_{k-1}^{t}(\theta)].
\end{equation}

\subsection{Marginal Cases}
The marginal cases for the walker-walker case produce trivial results; however, they provide us with some insight into the connection of OTOCs and localization. 
For $\theta=0$, $[U_k(0),U_{k'}(0)]=0$; $\forall k,k^{'}$, hence quite simply $F_{WW}(t)=0$ at all times.
For the case of  $\theta=\pi/2$; using  Eq.~(\ref{eq:u_k matrix for theta=pi/2}), results also in $U_{k-2}^{-t}(\theta)U_{k-1}^t (\theta)U_{k}^{-t}(\theta)U_{k-1}^{t}(\theta)=\mathds{1}$ and therefore $F_{WW}(t)=0$.
Thus, $F_{WW}(t)$ for both the marginal cases do not increase with time, even though the two marginal case  produce two different kinds of walk on the lattice, the information encoded in them remains localized in either one or two lattice sites for all time steps. The study of participation ratio which is a measure of localization also give similar results for these marginal cases \cite{omanakuttan2018quantum}, and we will comment on the connections to 
delocalization also in the Walker-coin case.
 \subsection{Non-Marginal Case}
\begin{figure*}[] 
    \subfloat{\includegraphics[width = 3.2in]{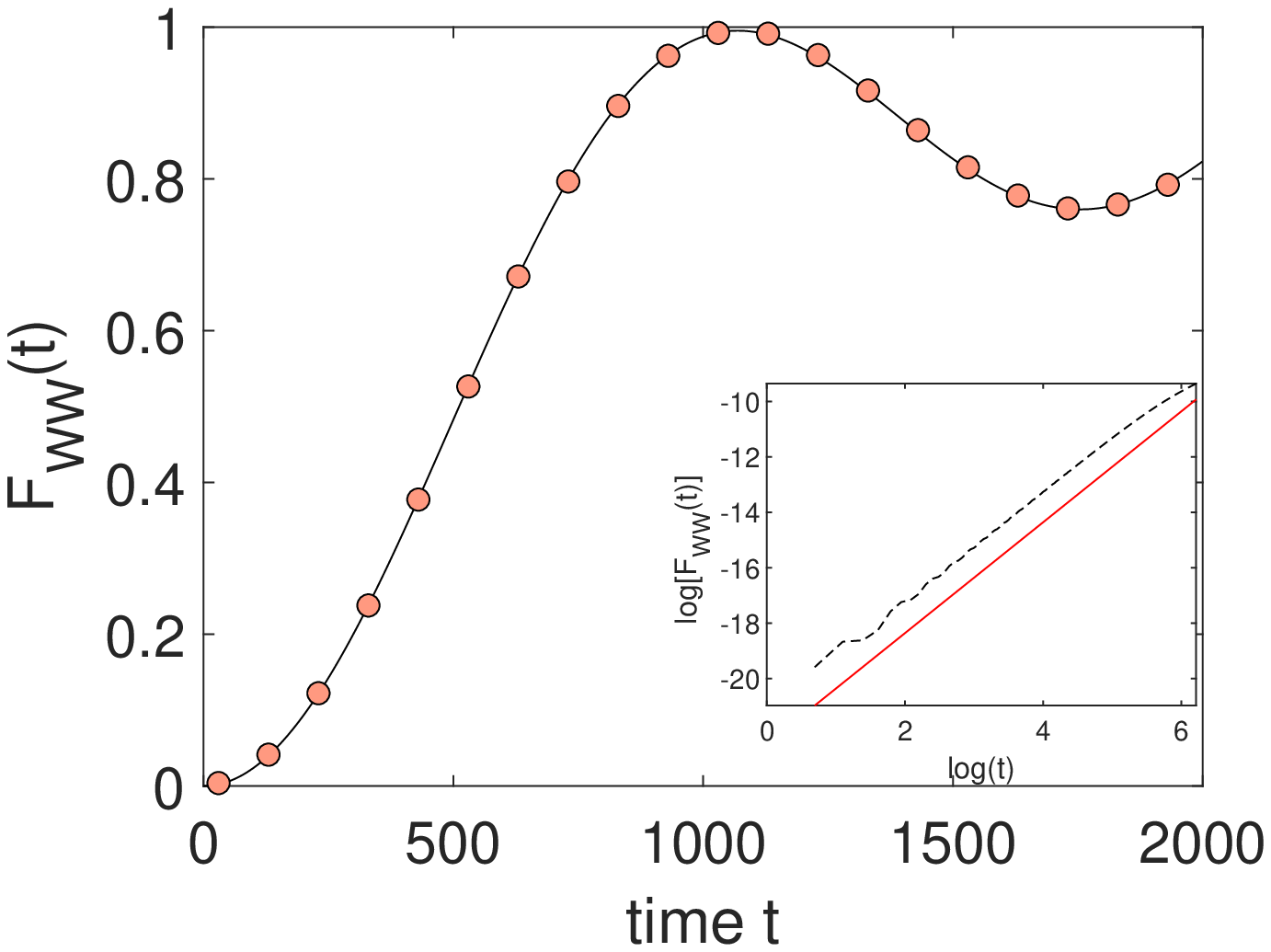}}  
   \subfloat{\includegraphics[width = 3.2in]{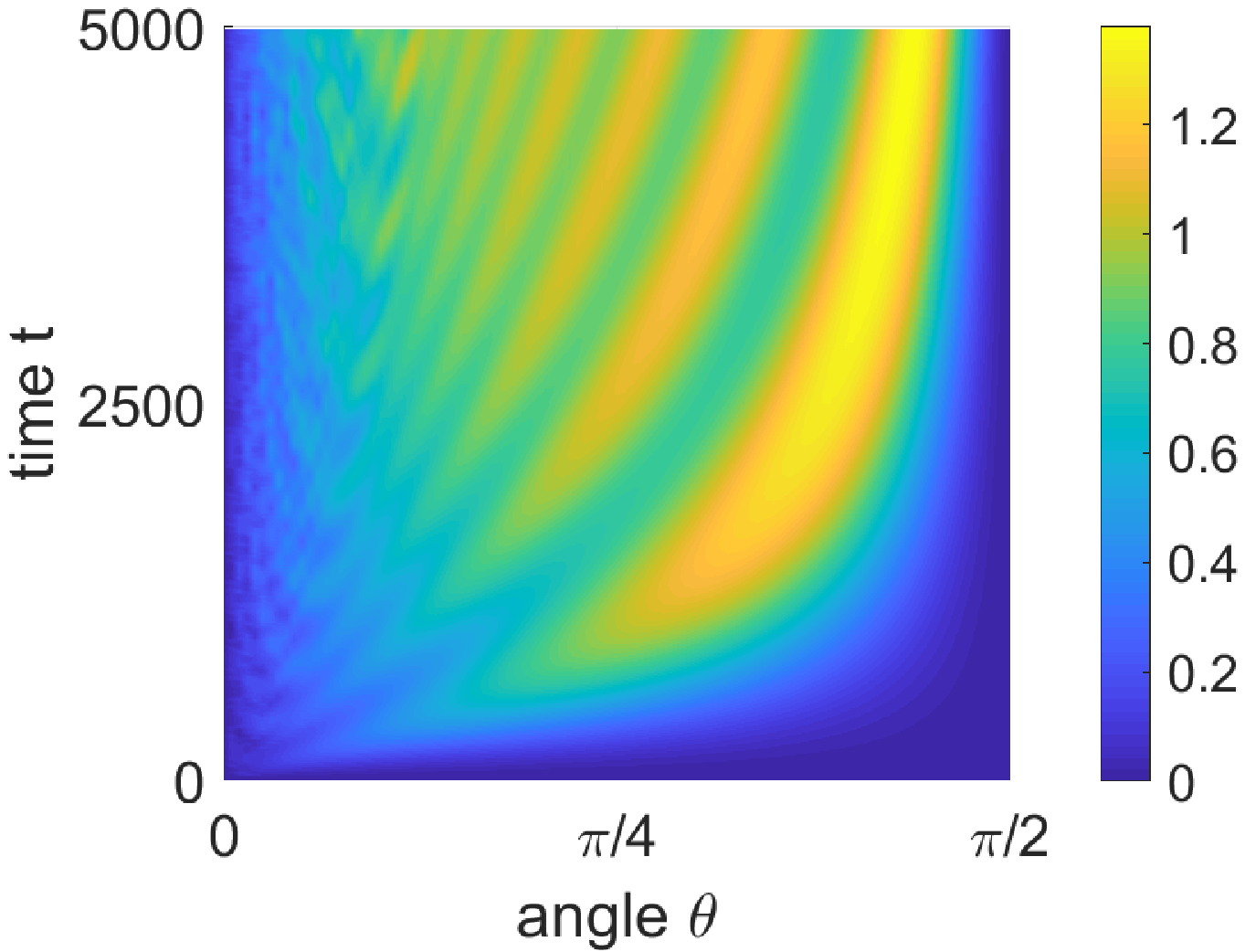}} 
 \caption{ Growth of $F_{WW}$ with time for the walker-walker case with $\theta=\pi/4$ is given on the left, black line is the $F_{CW}$ and the large dots are the one obtained using the numerical integration (plotted for time steps of $50$), the inset figure gives the $log-log$ graph of the same(dotted line) and the unbroken line is one with slope $2$ given for comparison  and the dependence of the $F_{WW}$ on time $t$ and angle $\theta$ is given on the right, for a lattice with size $N=100.$ }
 \label{fig:OTOC for the walker space Case for CSW}
 \end{figure*}

 Further simplification of $F_{WW}(t)$ in non-marginal cases is not easy. As an useful approximation, we consider the eigenvalues of $U_k$ to be different for different values of $k$, but we ignore the change in the eigenvectors,
 as $k$ changes by utmost $2$ units within each product in the sum of Eq.~(\ref{eq:OTOC for the walker-walker space }). This yields, 
  \begin{equation}
\label{eq:OTOC for the walker-walker space using eigenvalues 1}
C_4(t)=\frac{1}{2N}\sum_{l=\pm}\sum_{k=0}^{N-1}(\lambda_{k-2}^l\lambda_{k}^l)^t(\lambda_{k-1}^l)^{2t},
\end{equation}
where, 
\begin{equation}
\label{eq:eigenvalue of quantum walk}
\lambda^{\pm}_{k}= \exp(\pm i \alpha_k)
\end{equation}
are the eigenvalues of $U_k(\theta)$ and $\alpha_k$ is given by Eq.~(\ref{eq:Diagonalization of block matrix}).
This simplifies to,
 \begin{equation}
\label{eq:OTOC for the walker-walker space using eigenvalues 2}
C_4(t)=\frac{1}{N}\sum_{k=0}^{N-1} \cos\left[\left(2\alpha_{k-1}-\alpha_{k-2}-\alpha_{k}\right)t\right].
\end{equation}
Considering that $\alpha_k$ change slowly for large $N$, the second difference is approximated as the second
derivative 
\begin{equation}
\label{eq:seconddiffapprox}
\frac{d^2 \alpha_k}{dk^2}=-\frac{4\pi^2}{N^2} \frac{\sin \frac{2 \pi k}{N} \cos\theta\sin^2 \theta}{(1-\cos^2 \theta \sin ^2 \frac{2 \pi k}{N} )^{3/2}}.
\end{equation} 
Further approximating the sum as an integral yields 
 \begin{equation}
\begin{aligned}
\label{eq:OTOC in the walker-walker integral approximation}
C_4(t)\approx  \frac{1}{2\pi} \int_0^{2 \pi} \cos(\frac{4 \pi^2    t \cos \theta \sin^2 \theta \sin x }{N^2(1-\sin^2 x\cos^2 \theta)^{\frac{3}{2}}})dx.
\end{aligned}
\end{equation}
It is found that the numerical integration of the above is very close to the actual value $C_4(t)$ and is evident from the Fig.~(\ref{fig:OTOC for the walker space Case for CSW}). Now retaining only the first two terms  in the expansion of $\cos y$ and considering the case for $\theta=\pi/4$ in the above integral yields,
\begin{equation}
\label{eq:OTOC in the walker-walker approximation using cos}
F_{WW}(t)=1-C_4(t)\approx \frac{1}{2\sqrt{2}}\left(\frac{7 \pi^4 t^2}{ N^4 }\right).
\end{equation}
displaying the quadratic growth phase lasting a time $\sim N^2/7\pi^4$. 
%
Interestingly the integral approximation is close to the actual $F_{WW}$ despite the many approximations made.

In general for the non-marginal cases  the expression of $F_{WW}(t)$ is for short times approximately given as, 
\begin{equation}
\label{eq:F_ww for non-marginal}
F_{WW}(t)\approx\frac{\pi^4 t^2}{2 N^4} \frac{\cos ^2 \theta} {\sin \theta} \left(7 -\cos 2\theta\right).
\end{equation}
Hence the quadratic initial phase growth is valid for all values of $\theta$ except the marginal cases, up to a time scale approximately $\mathcal{O}(N^2).$ Also, from the above expression it is clear that growth rate is higher for $\theta$ close to zero compared to $\theta$ close to $\pi/2$.
The participation ratio, a measure of delocalization of the walker state, is correlated in the sense that it has comparitively larger value around $\theta=0$ that $\theta=\pi/2$ \cite{omanakuttan2018quantum}. The higher growth rate of $F_{WW}$ for $\theta=0$ could be considered as a consequence of this. In a somewhat different context, the OTOC also differentiates the many-body localized and thermal phases \cite{chen2017out}.

 The quadratic growth of $F_{WW}(t)$ implies that the standard discrete coined model of quantum walk discussed in this paper does not exhibit any quantum chaos in the walker space. From our previous work on the growth of Eherenfest time ($t_E$) with lattice dimension for discrete quantum walk ($N$) \cite{omanakuttan2018quantum}, it is found that $t_E \sim \sqrt{N}$. However, for systems exhibiting quantum chaos $t_E \sim \log{N}$ and hence the algebraic growth indicates lack of quantum chaos in the system which is also substantiated by the quadratic growth of $F_{WW}$ with time.

  \section{OTOC\smaller{s} for the coin-walker Case}
 The two cases considered above had operators from the same subspace, namely coin-coin or walker-walker, while OTOC is most interesting when studying cross correlations and 
one operator is in the coin space while the other is in the walker space. In this case OTOC indicates the scrambling of information that is localized in the coin to the walker and vice-versa. Consider the operator  $A=H\otimes \mathds{1}_N$ in the coin space  and $B=\mathds{1}_2\otimes \Tp$ in the walker space. Once again using the block diagonalization of  $U$  discussed in the previous sections we get the OTOC as 
  \begin{equation}
  \label{eq:OTOC for the walker coin case using block diagonal representation}  
  \begin{aligned}
  F_{CW}(t)=1-\frac{1}{2N} \sum_{k}\text{Tr}\left[U_{k-1}^{-t}(\theta)U_{k-2}^t(\theta) H U_{k-1}^{-t}(\theta)U_{k}^t(\theta)H\right].
  \end{aligned}
   \end{equation}
\subsection{Marginal Cases}  
For these cases further simplification of the Eq.~\eqref{eq:OTOC for the walker coin case using block diagonal representation} is possible and the expression of $F_{CW}(t)$ for $\theta=0$ or the $\sigma_z-$ coin is simply
\begin{equation}
\label{eq:coin-walker case for theta=0}
F_{CW}(t)=\sin^2 \left( \frac{2 \pi t}{N}\right),
\end{equation}
and for $\theta=\pi/2$ or the $\sigma_x-$ coin is,
\begin{equation}
F_{CW}(t)=\begin{cases}
0\,\, \text{if $t$ is even}\\
\sin^2 \left( \frac{2 \pi }{N}\right)\, \text{if $t$ is odd.}
\end{cases}
\label{eq:coin-walker case for theta=pi/2}
\end{equation}
The details of the calculation is given in the App.~(\ref{app:walker coin space}).
For $\sigma_x$-coin, the oscillations with time period $t=2$, is same as that of  the  coin-coin case counterpart given by Eq.~\eqref{eq:coin-walker case for theta=pi/2}; however, the amplitudes are not the same. The $\sigma_z$-coin shows a quadratic increase $\sim t^2/N^2$ remarkably different from the behavior of the counterpart for coin-coin, wherein $F$ attains a constant value after $t=1$ and changes very infrequently as given in Eq.~\eqref{eq:Coin-coin case for theta=0}

\subsection{Non marginal Cases}
\begin{figure*}[]
\subfloat{\includegraphics[width = 3.2in]{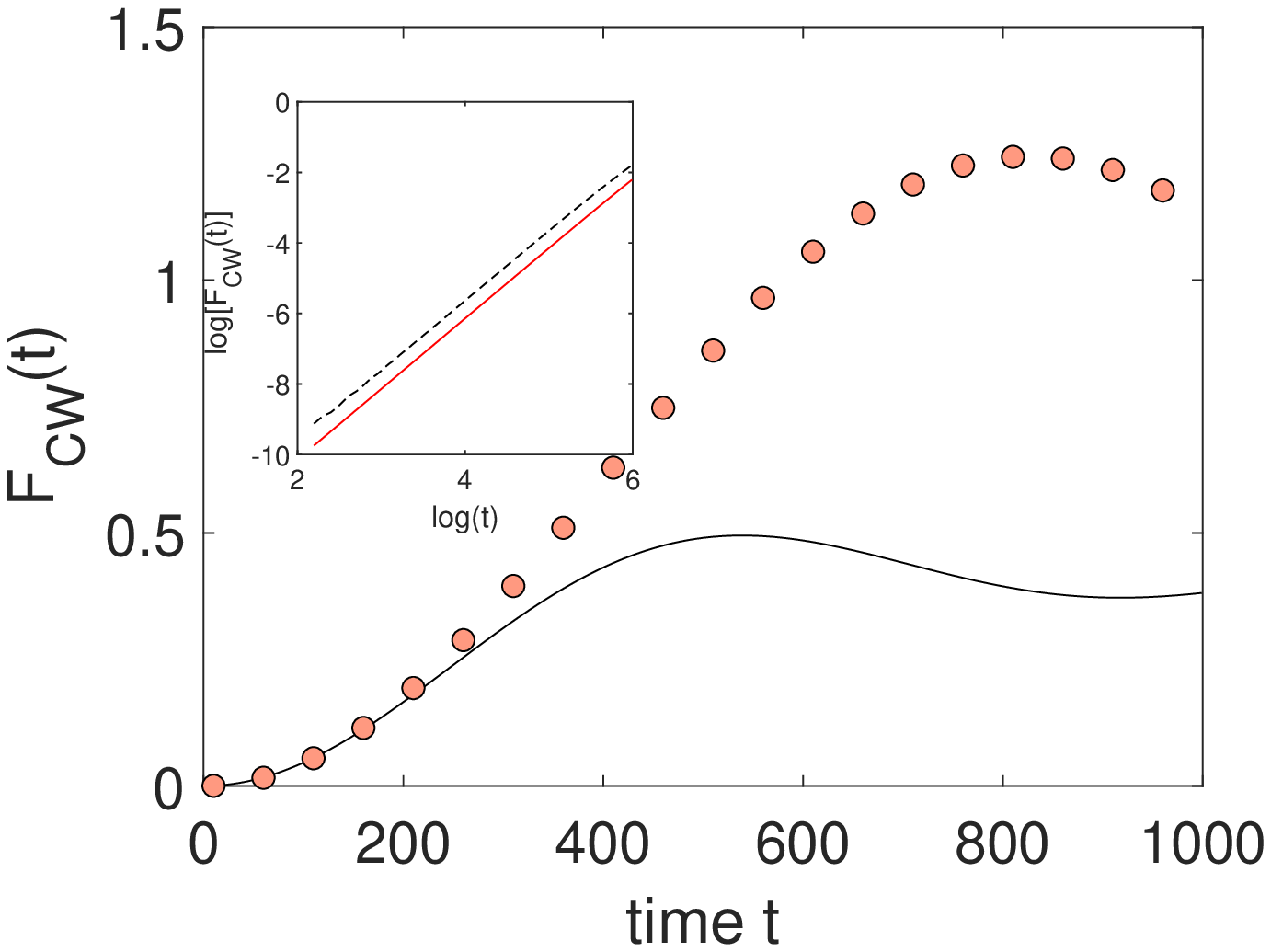}}
   \subfloat{\includegraphics[width = 3.2in]{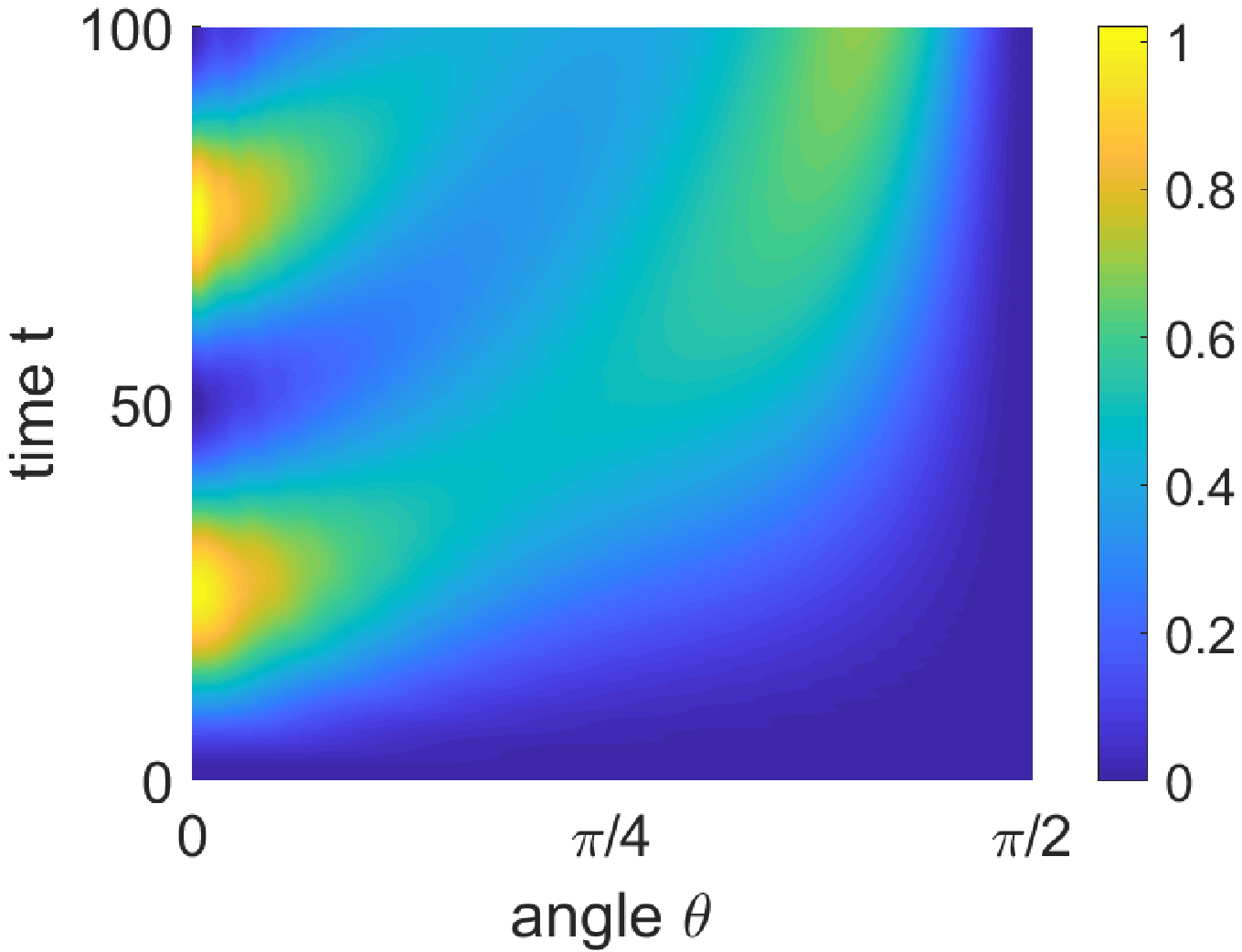}}
     \caption{The growth of $F_{CW}(t)$ with time for $\theta=\pi/4$, black line is the $F_{CW}$ and the large dots are the one obtained using the numerical integration (plotted for time steps of $50$), and the inset figure gives the $log-log$ graph of the same (dotted line) and the unbroken line is one with slope $2$ given for comparison for the coin-walker case with  a lattice size $N=1000$ (left) and the dependence of $F_{CW}(t)$ on time $t$t and angle $\theta$ (right) or a lattice of dimension $N=100$.  } 
 \label{fig:OTOC for the walker coin Case for CSW}
 \end{figure*}

In general, the four-time correlator is
  \begin{equation}
  C_4(t)=\sum_{k=0}^{N-1} \text{Tr}\left[U_{k-1}^{-t}(\theta)U_{k-2}^t(\theta) H U_{k-1}^{-t}(\theta)U_{k}^t(\theta)H\right].
\end{equation}
Again, further simplification of the above expression appears very hard and we need to use some approximation. Following the same routes as for the previous cases, we use the diagonal basis of the $U_k$, (Eq.\eqref{eq:Diagonalization of block matrix} and Eq.\eqref{eq:transformation of Hadamard matrix}) and ignore the change in the eigenvectors for each term in the sum as $k$ changes by utmost $2$. Combining these two approximations followed by some algebra yields,
  \begin{equation}
  \label{eq:OTOC for the walker coin approximation 2}
  \begin{aligned}
  C_4(t)\approx\frac{1}{N} \sum_{k=0}^{N-1}\left[a_{11}^2(k)\cos\left( 2t \frac{d\alpha_k}{dk} \right)+|a_{12}(k)|^2\cos \left( t \frac{d^2\alpha_k}{dk^2}\right)\right].
  \end{aligned}  
  \end{equation}
  The last approximation replaced first and second differences of the phases $\alpha_k$ by first and second derivatives, identical to the 
  approximation of  Eq.~(\ref{eq:seconddiffapprox}) used in simplifying 
Eq.~(\ref{eq:OTOC for the walker-walker space using eigenvalues 2}).
Substituting the values of $a_{11}(k)$ and $a_{12}(k)$ from Eq.~\eqref{eq:explicit expression of a_11} and Eq.~\eqref{eq:explicit expression of a_11 and a_12}, gives the integral approximation to the above sum for the special case of the Hadamard coin, $\theta=\pi/4$ as,
 \begin{equation}
\begin{aligned}
\label{eq:OTOC in the walker-coin integral approximation}
C_4(t)\approx &\frac{1}{2\pi}\left[ \int_0^{2 \pi}\frac{2 \sin^2 x}{3+ \cos 2x } \cos(\frac{\sqrt{2} \pi^2    t \sin x }{N^2(1-\frac{\sin^2 x}{2})^{\frac{3}{2}}})dx\right]\\+&\frac{1}{2\pi}\left[\int_0^{2 \pi}\frac{4 \cos^2 x}{3+ \cos 2x } \cos(\frac{\sqrt{2} \pi    t  \cos x }{N(1-\frac{\sin^2 x}{2})^{\frac{1}{2}}})dx\right].
\end{aligned}
\end{equation}
It is found that the numerical integration of above is close to the actual value $C_4(t)$ upto a time of order $\mathcal{O}(N/\pi)$ and is evident from the Fig.~(\ref{fig:OTOC for the walker coin Case for CSW}).
As 
\begin{equation}
\label{eq:OTOC in the walker-coin integral approximation 1}
F_{CW}(t)\approx \frac{0.591791}{4 \pi}\left(\frac{2\pi t}{N}\right)^2,
\end{equation}
the smaller time-scale indicates the time by which the coin information is scrambled and lost into the walker space. 
Thus this is a 
substantially long time and diverges as $N$ goes to $\infty$.
Therefore, the growth of $F_{CW}(t)$ is quadratic, however, the time scale of growth is of $\mathcal{O}(N)$ unlike $F_{WW}$ where it is of $\mathcal{O}(N^2)$. 

The growth of $F_{CW}(t)$ for coin-walker case is  given in Fig.~(\ref{fig:OTOC for the walker coin Case for CSW}). $F_{CW}(t)$ attains its maximum value around $\theta=0$ and its minimum around $\theta=\pi/2$. The behavior of the non-marginal cases are almost identical except the maximum value achieved. The time scale of growth of all these non-marginal cases are of the $\mathcal{O}(N).$
It maybe remarked that the growth of $F_{CW}(t)$ most closely resembles the growth of variance of the walker for in the coined quantum walk. Apart from the quadratic growth of both quantities, the nature of the marginal cases are similar, these cases being discussed in \cite{omanakuttan2018quantum}. It may be 
noted that the OTOC for projector observables are closely related to transition probabilities \cite{lakshminarayan2018out} and may provide insights into connections between various dynamical quantities.

\section{Summary and Conclusion}

This paper studied out-of time-ordered correlators for coined discrete quantum walks with a family of coin operators parametrized by an angle. We considered separate scenarios where both operators are in the coin-space, both are in the walker-space, or where one operator belonged to each space. Since the unitary operator corresponding to the quantum walk can be block-diagonalized in the momentum basis of the walker, the OTOCs were derived largely analytically. In the case of the marginal cases of the $\sigma_x$-coin and $\sigma_z$-coin complete solutions were possible. Primarily we found that in the coin-coin case OTOC
had a time scale of $O(1)$, beyond which it shows small fluctuations, whereas in the other two cases there was a quadrtic growth reminiscent of the variance
growth in the quantum walk. Here again the most interesting case is that of the coin-walker cross OTOC which explores the scrambling 
between the degrees of freedom of the coin and the walker. This has a characteristic time scale of $O(N)$ and shares the maximum similarity to the variance as well. It is clear that the dimensionalities of the spaces considered determined the time-scales of the OTOC and the largest was the scrambling of the walker within the walker space, which also grows quadratically and for a time $O(N^2)$. 

While the study reflects the lack of quantum chaos in the quantum walk, it also opens the way to analyze various other walks such as higher-dimensional cases and a complete family of coin-operators defined by $SU(2)$ coins. There is a class of quantum walk defined in \cite{lakshminarayan2003random} where the coin dynamics is governed by operators whose classical limits are chaotic, wherein the coin-coin OTOC will have an exponential growth in time. How this coin-scrambling affects the walker 
will be the subject of future work. We hope that our exploratory analysis of the OTOC in quantum walks enables further characterization 
and elucidation of the walks themselves, while at the same time add meaningfully to the burgeoning literature on the OTOC in a context not yet studied.
\section{Acknowledgment}
 SO would like to thank Gopikrishnan Muraleedharan (CQuIC, University of New Mexico, Albuquerque) for his help during various stages of the progress of the work.

\appendix
\section{ $F_{CC}$ for the  Marginal cases }
\label{app:coin space}
\subsection{Case I: $\theta=0$}
$F_{CC}$ is given in terms of block diagonal matrices in the Eq.~(\ref{eq:OTOC in the Coin-Coin Case using block diagonal matrix}). Now for $\theta=0$ and $A=H\otimes\mathds{1}$,
\begin{equation}
\label{eq:u_k matrix for theta=0}
U^t_{k}=\begin{cases}\begin{pmatrix}
\omega^{kt} & 0\\
0 & -\omega^{-kt}
\end{pmatrix}\text{ if $t$ is odd}\\
\begin{pmatrix}
\omega^{kt} & 0\\
0 & \omega^{-kt}
\end{pmatrix} \text{if $t$ is even}.
\end{cases}
\end{equation}
 Hence Eq.~\eqref{eq:OTOC in the Coin-Coin Case using block diagonal matrix}  becomes,
\begin{equation}
C_4(t)=\frac{1}{4}\left(-1+4(-1)^t\sum_{k=0}^{N-1}\cos \frac{4 \pi k t}{N}+ \sum_{k=0}^{N-1} \cos \frac{8 \pi k t}{N}\right)
\end{equation}
which can be further simplified as,
\begin{equation}
C_4(t)=\frac{1}{4}\left[-1+4(-1)^t\, \delta(2 t\,\text{mod}N,0)+\delta(4 t\,\text{mod}N,0)\right],
\end{equation}
and thus,
\begin{equation}
F_{CC}(t)=\frac{5}{4}-\frac{1}{4}\left[4(-1)^t \, \delta(2 t\,\text{mod}N,0)+\delta(4 t\,\text{mod}N,0)\right].
\end{equation}
\subsection{Case II: $\theta=\frac{\pi}{2}$}
For $\theta=\frac{\pi}{2},$ following some trivial algebra,
 \begin{equation}
 \label{eq:u_k matrix for theta=pi/2}
 U^{t}_k=U^{-t}_k=\begin{cases}
 \begin{pmatrix}
 1 & 0\\
 0 & 1
 \end{pmatrix}\, \text{if $t$ is even}\\
 \begin{pmatrix}
 0 & \omega^{k}\\
 \omega^{-k} & 0
 \end{pmatrix} \,\text{for $t$ odd,}
 \end{cases}
 \end{equation}
 which in turn gives,
 \begin{equation}
U^{-t}_{k}H U^t_{k}H=\begin{cases}
\begin{pmatrix}
1 & 0\\
0 & 1
\end{pmatrix}\, \text{if $t$ is even,}\\
\frac{1}{2}\begin{pmatrix}
-1+\omega^{2k} & -1-\omega^{2k}\\
1+\omega^{-2k} & -1+\omega^{2k} 
\end{pmatrix}\,\text{for $t$ odd.}
\end{cases}
\end{equation}
Therefore;
\begin{equation}
C_4(t)=\begin{cases}
1\,\, \text{if $t$ is even, }\\
-\frac{1}{4}\, \text{for $t$ odd,}
\end{cases}
\end{equation}
and hence,
\begin{equation}
F_{CC}(t)=\begin{cases}
0\,\, \text{if $t$ is even, }\\
\frac{5}{4}\, \text{for $t$ odd.}
\end{cases}
\end{equation}
\section{Explicit Calculation of $a_{11}(k)$ and $a_{12}(k)$ for $\theta=\frac{\pi}{4}$}
\label{app:Explicit Calculation of a11}
Using Eq.~\eqref{eq:block diagonal representation of U_csw} for $\theta=\pi/4$ and substituting $y_k=\exp(2\pi i k/N)$ yields,
\begin{equation}
\label{eq:block diagonal matrix for U_csw hadamard}
U_k\left(\frac{\pi}{4}\right)=\frac{1}{\sqrt{2}}\begin{pmatrix}
y_k & y_k\\
\frac{1}{y_k} &-\frac{1}{y_k} 
\end{pmatrix}.
\end{equation}
The eigenvectors of which are given as: 
\begin{equation}
\label{eq:eigenvector of Hadamard }
\begin{pmatrix}
\frac{1}{2}(1 + y_k^2 \pm \sqrt{1 + 6 y_k^2 + y_k^4}),  &1
\end{pmatrix},
\end{equation}
which can be further simplified and  $\T_k^{\dagger}$ can be written as, 
\begin{equation}
\label{eq:eigenvector matrix of Hadamard }
T_k^{\dagger}=\begin{bmatrix}
C_{-}(x) \left(\cos[x] - \sqrt{1 + \cos[x]^2} , \exp -i x\right)\\ 
 C_{+}(x) \left(\cos[x] +\sqrt{1 + \cos[x]^2}, \exp -i x\right)
\end{bmatrix}^T,
\end{equation}
where $x=2\pi k/N$ and the normalization constants $C_{\pm}$ are given as 
\begin{equation}
\label{eq:Normalization constants}
C_{\pm}(x)=\left(2\sqrt{[1 + \cos^2 x][1 + \cos^2 x\pm\cos x]}\right)^{-\frac{1}{2}}.\\
\end{equation}
Now  $T_k H T_k^{\dagger}$ can be calculated and following some  algebra it is found that,
\begin{equation}
\label{eq:explicit expression of a_11}
a_{11}(k)=-\frac{2 \cos x}{\sqrt{3 + \cos 2 x}},
\end{equation}
and hence, 
\begin{equation}
\label{eq:explicit expression of a_11 and a_12}
\begin{aligned}
a_{11}^2(k)=&\frac{4 \cos^2 x} {3 + \cos 2 x},\\
|a_{12}(k)|^2=&\frac{2 \sin^2 x} {3 + \cos 2 x}.
\end{aligned}
\end{equation}
It is thus possible to calculate the average values,
\begin{equation}
\label{eq:Average values of a11}
\begin{aligned}
\left<a_{11}^4(k)\right>=&4-\frac{5}{\sqrt{2}},\\
\left<|a_{12}(k)|^4\right>=&1-\frac{1}{\sqrt{2}},\\
\left<a_{11}^2(k)|a_{12}(k)|^2\right>=&-2+\frac{3}{\sqrt{2}}.
\end{aligned}
\end{equation}

\section{ OTOCs for the two level system}
\label{app:two level atom}
Consider the Hamiltonian defined as, 
\begin{equation}
\label{eq:Hamiltonian of the two level system}
H=\cos \theta \sigma_x+ \sin \theta \sigma_y,
\end{equation}
and say $A=B=\sigma_z$, therefore 
\begin{equation}
\label{eq:Time evolution in two level atom}
A(t)=\begin{pmatrix}
\cos 2t &-i \sin 2t\, e^{-i\theta}\\
i\sin 2t\, e^{i\theta} & -\cos 2t
\end{pmatrix},
\end{equation}
hence $C_4(t)=\cos 4t$, which gives,
\begin{equation}
F_{}(t)=1-\cos 4t.
\end{equation}

\section{ $F_{CW}$ for the  Marginal cases}

\label{app:walker coin space}

\subsection{Case I: $\theta=0$}
 $F_{CW}$ is given in terms of block diagonal matrices in Eq.~(\ref{eq:OTOC for the walker coin case using block diagonal representation}). Let $T=U_{k-1}^{-t}(\theta)U_{k-2}^t(\theta) H U_{k-1}^{-t}(\theta)U_{k}^t(\theta)H$ and using Eq.~(\ref{eq:u_k matrix for theta=0}) for $\theta=0$,
\begin{equation}
T=\frac{1}{2}\begin{pmatrix}
1+\omega^{-2t} &1-\omega^{-2t}\\
-1+\omega^{-2t} &1+\omega^{2t}
\end{pmatrix}.
\end{equation}
which gives $C_4(t)=\frac{1}{2}(1+\cos \frac{4 \pi t}{N}),$ and hence
\begin{equation}
F_{CW}(t)= \sin^2 \left( \frac{2 \pi t}{N}\right).
\end{equation}
\subsection{Case II: $\theta=\frac{\pi}{2}$}
Using Eq.~(\ref{eq:u_k matrix for theta=pi/2}) for $\theta=\pi/2$ yields,
\begin{equation}
T=\begin{cases}
\begin{pmatrix}
1 &0\\
0 & 1
\end{pmatrix}\, \text {if $t $ is even }\\
\frac{1}{2}\begin{pmatrix}
1+\omega^{2} &1-\omega^{2}\\
-1+\omega^{-2} &1+\omega^{-2}
\end{pmatrix}\, \text{if $t$ is odd}.
\end{cases}
\end{equation}
which in turn gives,
\begin{equation}
C_4(t)=\begin{cases}
1\,\, \text{if $t$ is even}\\
\frac{1}{2}(1+\cos \frac{4 \pi }{N})\, \text{if $t$ is odd,}
\end{cases}
\end{equation}
therefore,
\begin{equation}
F_{CW}(t)=\begin{cases}
0\,\, \text{if $t$ is even}\\
\frac{1}{2}(1-\cos \frac{4 \pi }{N})\, \text{if $t$ is odd.}
\end{cases}
\end{equation}

\bibliography{otocreference}

\begin{thebibliography}{40}%
\makeatletter
\providecommand \@ifxundefined [1]{%
 \@ifx{#1\undefined}
}%
\providecommand \@ifnum [1]{%
 \ifnum #1\expandafter \@firstoftwo
 \else \expandafter \@secondoftwo
 \fi
}%
\providecommand \@ifx [1]{%
 \ifx #1\expandafter \@firstoftwo
 \else \expandafter \@secondoftwo
 \fi
}%
\providecommand \natexlab [1]{#1}%
\providecommand \enquote  [1]{``#1''}%
\providecommand \bibnamefont  [1]{#1}%
\providecommand \bibfnamefont [1]{#1}%
\providecommand \citenamefont [1]{#1}%
\providecommand \href@noop [0]{\@secondoftwo}%
\providecommand \href [0]{\begingroup \@sanitize@url \@href}%
\providecommand \@href[1]{\@@startlink{#1}\@@href}%
\providecommand \@@href[1]{\endgroup#1\@@endlink}%
\providecommand \@sanitize@url [0]{\catcode `\\12\catcode `\$12\catcode
  `\&12\catcode `\#12\catcode `\^12\catcode `\_12\catcode `\%12\relax}%
\providecommand \@@startlink[1]{}%
\providecommand \@@endlink[0]{}%
\providecommand \url  [0]{\begingroup\@sanitize@url \@url }%
\providecommand \@url [1]{\endgroup\@href {#1}{\urlprefix }}%
\providecommand \urlprefix  [0]{URL }%
\providecommand \Eprint [0]{\href }%
\providecommand \doibase [0]{http://dx.doi.org/}%
\providecommand \selectlanguage [0]{\@gobble}%
\providecommand \bibinfo  [0]{\@secondoftwo}%
\providecommand \bibfield  [0]{\@secondoftwo}%
\providecommand \translation [1]{[#1]}%
\providecommand \BibitemOpen [0]{}%
\providecommand \bibitemStop [0]{}%
\providecommand \bibitemNoStop [0]{.\EOS\space}%
\providecommand \EOS [0]{\spacefactor3000\relax}%
\providecommand \BibitemShut  [1]{\csname bibitem#1\endcsname}%
\let\auto@bib@innerbib\@empty
\bibitem [{\citenamefont {Larkin}\ and\ \citenamefont
  {Ovchinnikov}(1969)}]{larkin}%
  \BibitemOpen
  \bibfield  {author} {\bibinfo {author} {\bibfnamefont {A.}~\bibnamefont
  {Larkin}}\ and\ \bibinfo {author} {\bibfnamefont {Y.~N.}\ \bibnamefont
  {Ovchinnikov}},\ }\href@noop {} {\bibfield  {journal} {\bibinfo  {journal}
  {Soviet Journal of Experimental and Theoretical Physics}\ }\textbf {\bibinfo
  {volume} {28}},\ \bibinfo {pages} {1200} (\bibinfo {year}
  {1969})}\BibitemShut {NoStop}%
\bibitem [{\citenamefont {Maldacena}\ \emph {et~al.}(2016)\citenamefont
  {Maldacena}, \citenamefont {Shenker},\ and\ \citenamefont
  {Stanford}}]{maldacena2016bound}%
  \BibitemOpen
  \bibfield  {author} {\bibinfo {author} {\bibfnamefont {J.}~\bibnamefont
  {Maldacena}}, \bibinfo {author} {\bibfnamefont {S.~H.}\ \bibnamefont
  {Shenker}}, \ and\ \bibinfo {author} {\bibfnamefont {D.}~\bibnamefont
  {Stanford}},\ }\href@noop {} {\bibfield  {journal} {\bibinfo  {journal}
  {Journal of High Energy Physics}\ }\textbf {\bibinfo {volume} {2016}},\
  \bibinfo {pages} {106} (\bibinfo {year} {2016})}\BibitemShut {NoStop}%
\bibitem [{\citenamefont {Moudgalya}\ \emph {et~al.}(2018)\citenamefont
  {Moudgalya}, \citenamefont {Devakul}, \citenamefont {von Keyserlingk},\ and\
  \citenamefont {Sondhi}}]{moudgalya2018operator}%
  \BibitemOpen
  \bibfield  {author} {\bibinfo {author} {\bibfnamefont {S.}~\bibnamefont
  {Moudgalya}}, \bibinfo {author} {\bibfnamefont {T.}~\bibnamefont {Devakul}},
  \bibinfo {author} {\bibfnamefont {C.}~\bibnamefont {von Keyserlingk}}, \ and\
  \bibinfo {author} {\bibfnamefont {S.}~\bibnamefont {Sondhi}},\ }\href@noop {}
  {\bibfield  {journal} {\bibinfo  {journal} {arXiv preprint arXiv:1808.04889}\
  } (\bibinfo {year} {2018})}\BibitemShut {NoStop}%
\bibitem [{\citenamefont {Shenker}\ and\ \citenamefont
  {Stanford}(2014)}]{shenker2014black}%
  \BibitemOpen
  \bibfield  {author} {\bibinfo {author} {\bibfnamefont {S.~H.}\ \bibnamefont
  {Shenker}}\ and\ \bibinfo {author} {\bibfnamefont {D.}~\bibnamefont
  {Stanford}},\ }\href@noop {} {\bibfield  {journal} {\bibinfo  {journal}
  {Journal of High Energy Physics}\ }\textbf {\bibinfo {volume} {2014}},\
  \bibinfo {pages} {67} (\bibinfo {year} {2014})}\BibitemShut {NoStop}%
\bibitem [{\citenamefont {Polchinski}\ and\ \citenamefont
  {Rosenhaus}(2016)}]{polchinski2016spectrum}%
  \BibitemOpen
  \bibfield  {author} {\bibinfo {author} {\bibfnamefont {J.}~\bibnamefont
  {Polchinski}}\ and\ \bibinfo {author} {\bibfnamefont {V.}~\bibnamefont
  {Rosenhaus}},\ }\href@noop {} {\bibfield  {journal} {\bibinfo  {journal}
  {Journal of High Energy Physics}\ }\textbf {\bibinfo {volume} {2016}},\
  \bibinfo {pages} {1} (\bibinfo {year} {2016})}\BibitemShut {NoStop}%
\bibitem [{\citenamefont {Roberts}\ \emph {et~al.}(2015)\citenamefont
  {Roberts}, \citenamefont {Stanford},\ and\ \citenamefont
  {Susskind}}]{roberts2015localized}%
  \BibitemOpen
  \bibfield  {author} {\bibinfo {author} {\bibfnamefont {D.~A.}\ \bibnamefont
  {Roberts}}, \bibinfo {author} {\bibfnamefont {D.}~\bibnamefont {Stanford}}, \
  and\ \bibinfo {author} {\bibfnamefont {L.}~\bibnamefont {Susskind}},\
  }\href@noop {} {\bibfield  {journal} {\bibinfo  {journal} {Journal of High
  Energy Physics}\ }\textbf {\bibinfo {volume} {2015}},\ \bibinfo {pages} {51}
  (\bibinfo {year} {2015})}\BibitemShut {NoStop}%
\bibitem [{\citenamefont {Shenker}\ and\ \citenamefont
  {Stanford}(2015)}]{shenker2015stringy}%
  \BibitemOpen
  \bibfield  {author} {\bibinfo {author} {\bibfnamefont {S.~H.}\ \bibnamefont
  {Shenker}}\ and\ \bibinfo {author} {\bibfnamefont {D.}~\bibnamefont
  {Stanford}},\ }\href@noop {} {\bibfield  {journal} {\bibinfo  {journal}
  {Journal of High Energy Physics}\ }\textbf {\bibinfo {volume} {2015}},\
  \bibinfo {pages} {132} (\bibinfo {year} {2015})}\BibitemShut {NoStop}%
\bibitem [{\citenamefont {Swingle}\ \emph {et~al.}(2016)\citenamefont
  {Swingle}, \citenamefont {Bentsen}, \citenamefont {Schleier-Smith},\ and\
  \citenamefont {Hayden}}]{swingle2016measuring}%
  \BibitemOpen
  \bibfield  {author} {\bibinfo {author} {\bibfnamefont {B.}~\bibnamefont
  {Swingle}}, \bibinfo {author} {\bibfnamefont {G.}~\bibnamefont {Bentsen}},
  \bibinfo {author} {\bibfnamefont {M.}~\bibnamefont {Schleier-Smith}}, \ and\
  \bibinfo {author} {\bibfnamefont {P.}~\bibnamefont {Hayden}},\ }\href@noop {}
  {\bibfield  {journal} {\bibinfo  {journal} {Physical Review A}\ }\textbf
  {\bibinfo {volume} {94}},\ \bibinfo {pages} {040302} (\bibinfo {year}
  {2016})}\BibitemShut {NoStop}%
\bibitem [{\citenamefont {Hashimoto}\ \emph {et~al.}(2017)\citenamefont
  {Hashimoto}, \citenamefont {Murata},\ and\ \citenamefont
  {Yoshii}}]{hashimoto2017out}%
  \BibitemOpen
  \bibfield  {author} {\bibinfo {author} {\bibfnamefont {K.}~\bibnamefont
  {Hashimoto}}, \bibinfo {author} {\bibfnamefont {K.}~\bibnamefont {Murata}}, \
  and\ \bibinfo {author} {\bibfnamefont {R.}~\bibnamefont {Yoshii}},\
  }\href@noop {} {\bibfield  {journal} {\bibinfo  {journal} {Journal of High
  Energy Physics}\ }\textbf {\bibinfo {volume} {2017}},\ \bibinfo {pages} {138}
  (\bibinfo {year} {2017})}\BibitemShut {NoStop}%
\bibitem [{\citenamefont {Syzranov}\ \emph {et~al.}(2018)\citenamefont
  {Syzranov}, \citenamefont {Gorshkov},\ and\ \citenamefont
  {Galitski}}]{PhysRevB.97.161114}%
  \BibitemOpen
  \bibfield  {author} {\bibinfo {author} {\bibfnamefont {S.~V.}\ \bibnamefont
  {Syzranov}}, \bibinfo {author} {\bibfnamefont {A.~V.}\ \bibnamefont
  {Gorshkov}}, \ and\ \bibinfo {author} {\bibfnamefont {V.}~\bibnamefont
  {Galitski}},\ }\href@noop {} {\bibfield  {journal} {\bibinfo  {journal}
  {Phys. Rev. B}\ }\textbf {\bibinfo {volume} {97}},\ \bibinfo {pages} {161114}
  (\bibinfo {year} {2018})}\BibitemShut {NoStop}%
\bibitem [{\citenamefont {Tsuji}\ \emph {et~al.}(2017)\citenamefont {Tsuji},
  \citenamefont {Werner},\ and\ \citenamefont {Ueda}}]{PhysRevA.95.011601}%
  \BibitemOpen
  \bibfield  {author} {\bibinfo {author} {\bibfnamefont {N.}~\bibnamefont
  {Tsuji}}, \bibinfo {author} {\bibfnamefont {P.}~\bibnamefont {Werner}}, \
  and\ \bibinfo {author} {\bibfnamefont {M.}~\bibnamefont {Ueda}},\ }\href@noop
  {} {\bibfield  {journal} {\bibinfo  {journal} {Phys. Rev. A}\ }\textbf
  {\bibinfo {volume} {95}},\ \bibinfo {pages} {011601} (\bibinfo {year}
  {2017})}\BibitemShut {NoStop}%
\bibitem [{\citenamefont {Kitaev}\ and\ \citenamefont
  {Suh}(2018)}]{kitaev2018soft}%
  \BibitemOpen
  \bibfield  {author} {\bibinfo {author} {\bibfnamefont {A.}~\bibnamefont
  {Kitaev}}\ and\ \bibinfo {author} {\bibfnamefont {S.~J.}\ \bibnamefont
  {Suh}},\ }\href@noop {} {\bibfield  {journal} {\bibinfo  {journal} {Journal
  of High Energy Physics}\ }\textbf {\bibinfo {volume} {2018}},\ \bibinfo
  {pages} {183} (\bibinfo {year} {2018})}\BibitemShut {NoStop}%
\bibitem [{\citenamefont {Lakshminarayan}(2019)}]{lakshminarayan2018out}%
  \BibitemOpen
  \bibfield  {author} {\bibinfo {author} {\bibfnamefont {A.}~\bibnamefont
  {Lakshminarayan}},\ }\href {\doibase 10.1103/PhysRevE.99.012201} {\bibfield
  {journal} {\bibinfo  {journal} {Phys. Rev. E}\ }\textbf {\bibinfo {volume}
  {99}},\ \bibinfo {pages} {012201} (\bibinfo {year} {2019})}\BibitemShut
  {NoStop}%
\bibitem [{\citenamefont {Seshadri}\ \emph {et~al.}(2018)\citenamefont
  {Seshadri}, \citenamefont {Madhok},\ and\ \citenamefont
  {Lakshminarayan}}]{PhysRevE.98.052205}%
  \BibitemOpen
  \bibfield  {author} {\bibinfo {author} {\bibfnamefont {A.}~\bibnamefont
  {Seshadri}}, \bibinfo {author} {\bibfnamefont {V.}~\bibnamefont {Madhok}}, \
  and\ \bibinfo {author} {\bibfnamefont {A.}~\bibnamefont {Lakshminarayan}},\
  }\href {\doibase 10.1103/PhysRevE.98.052205} {\bibfield  {journal} {\bibinfo
  {journal} {Phys. Rev. E}\ }\textbf {\bibinfo {volume} {98}},\ \bibinfo
  {pages} {052205} (\bibinfo {year} {2018})}\BibitemShut {NoStop}%
\bibitem [{\citenamefont {Aleiner}\ \emph {et~al.}(2016)\citenamefont
  {Aleiner}, \citenamefont {Faoro},\ and\ \citenamefont
  {Ioffe}}]{aleiner2016microscopic}%
  \BibitemOpen
  \bibfield  {author} {\bibinfo {author} {\bibfnamefont {I.~L.}\ \bibnamefont
  {Aleiner}}, \bibinfo {author} {\bibfnamefont {L.}~\bibnamefont {Faoro}}, \
  and\ \bibinfo {author} {\bibfnamefont {L.~B.}\ \bibnamefont {Ioffe}},\
  }\href@noop {} {\bibfield  {journal} {\bibinfo  {journal} {Annals of
  Physics}\ }\textbf {\bibinfo {volume} {375}},\ \bibinfo {pages} {378}
  (\bibinfo {year} {2016})}\BibitemShut {NoStop}%
\bibitem [{\citenamefont {Roberts}\ and\ \citenamefont
  {Stanford}(2014)}]{roberts2014two}%
  \BibitemOpen
  \bibfield  {author} {\bibinfo {author} {\bibfnamefont {D.~A.}\ \bibnamefont
  {Roberts}}\ and\ \bibinfo {author} {\bibfnamefont {D.}~\bibnamefont
  {Stanford}},\ }\href@noop {} {\bibfield  {journal} {\bibinfo  {journal}
  {arXiv preprint arXiv:1412.5123}\ } (\bibinfo {year} {2014})}\BibitemShut
  {NoStop}%
\bibitem [{\citenamefont {Aharonov}\ \emph {et~al.}(1993)\citenamefont
  {Aharonov}, \citenamefont {Davidovich},\ and\ \citenamefont
  {Zagury}}]{aharonov1993quantum}%
  \BibitemOpen
  \bibfield  {author} {\bibinfo {author} {\bibfnamefont {Y.}~\bibnamefont
  {Aharonov}}, \bibinfo {author} {\bibfnamefont {L.}~\bibnamefont
  {Davidovich}}, \ and\ \bibinfo {author} {\bibfnamefont {N.}~\bibnamefont
  {Zagury}},\ }\href@noop {} {\bibfield  {journal} {\bibinfo  {journal}
  {Physical Review A}\ }\textbf {\bibinfo {volume} {48}},\ \bibinfo {pages}
  {1687} (\bibinfo {year} {1993})}\BibitemShut {NoStop}%
\bibitem [{\citenamefont {Kempe}(2003)}]{kempe2003quantum}%
  \BibitemOpen
  \bibfield  {author} {\bibinfo {author} {\bibfnamefont {J.}~\bibnamefont
  {Kempe}},\ }\href@noop {} {\bibfield  {journal} {\bibinfo  {journal}
  {Contemporary Physics}\ }\textbf {\bibinfo {volume} {44}},\ \bibinfo {pages}
  {307} (\bibinfo {year} {2003})}\BibitemShut {NoStop}%
\bibitem [{\citenamefont {Childs}\ \emph {et~al.}(2002)\citenamefont {Childs},
  \citenamefont {Farhi},\ and\ \citenamefont {Gutmann}}]{childs2002example}%
  \BibitemOpen
  \bibfield  {author} {\bibinfo {author} {\bibfnamefont {A.~M.}\ \bibnamefont
  {Childs}}, \bibinfo {author} {\bibfnamefont {E.}~\bibnamefont {Farhi}}, \
  and\ \bibinfo {author} {\bibfnamefont {S.}~\bibnamefont {Gutmann}},\
  }\href@noop {} {\bibfield  {journal} {\bibinfo  {journal} {Quantum
  Information Processing}\ }\textbf {\bibinfo {volume} {1}},\ \bibinfo {pages}
  {35} (\bibinfo {year} {2002})}\BibitemShut {NoStop}%
\bibitem [{\citenamefont {Childs}(2009)}]{childs2009universal}%
  \BibitemOpen
  \bibfield  {author} {\bibinfo {author} {\bibfnamefont {A.~M.}\ \bibnamefont
  {Childs}},\ }\href@noop {} {\bibfield  {journal} {\bibinfo  {journal}
  {Physical review letters}\ }\textbf {\bibinfo {volume} {102}},\ \bibinfo
  {pages} {180501} (\bibinfo {year} {2009})}\BibitemShut {NoStop}%
\bibitem [{\citenamefont {Lovett}\ \emph {et~al.}(2010)\citenamefont {Lovett},
  \citenamefont {Cooper}, \citenamefont {Everitt}, \citenamefont {Trevers},\
  and\ \citenamefont {Kendon}}]{lovett2010universal}%
  \BibitemOpen
  \bibfield  {author} {\bibinfo {author} {\bibfnamefont {N.~B.}\ \bibnamefont
  {Lovett}}, \bibinfo {author} {\bibfnamefont {S.}~\bibnamefont {Cooper}},
  \bibinfo {author} {\bibfnamefont {M.}~\bibnamefont {Everitt}}, \bibinfo
  {author} {\bibfnamefont {M.}~\bibnamefont {Trevers}}, \ and\ \bibinfo
  {author} {\bibfnamefont {V.}~\bibnamefont {Kendon}},\ }\href@noop {}
  {\bibfield  {journal} {\bibinfo  {journal} {Physical Review A}\ }\textbf
  {\bibinfo {volume} {81}},\ \bibinfo {pages} {042330} (\bibinfo {year}
  {2010})}\BibitemShut {NoStop}%
\bibitem [{\citenamefont {Childs}\ and\ \citenamefont
  {Goldstone}(2004)}]{childs2004spatial}%
  \BibitemOpen
  \bibfield  {author} {\bibinfo {author} {\bibfnamefont {A.~M.}\ \bibnamefont
  {Childs}}\ and\ \bibinfo {author} {\bibfnamefont {J.}~\bibnamefont
  {Goldstone}},\ }\href@noop {} {\bibfield  {journal} {\bibinfo  {journal}
  {Physical Review A}\ }\textbf {\bibinfo {volume} {70}},\ \bibinfo {pages}
  {022314} (\bibinfo {year} {2004})}\BibitemShut {NoStop}%
\bibitem [{\citenamefont {Aaronson}\ and\ \citenamefont
  {Ambainis}(2003)}]{aaronson2003quantum}%
  \BibitemOpen
  \bibfield  {author} {\bibinfo {author} {\bibfnamefont {S.}~\bibnamefont
  {Aaronson}}\ and\ \bibinfo {author} {\bibfnamefont {A.}~\bibnamefont
  {Ambainis}},\ }in\ \href@noop {} {\emph {\bibinfo {booktitle} {Foundations of
  Computer Science, 2003. Proceedings. 44th Annual IEEE Symposium on}}}\
  (\bibinfo {organization} {IEEE},\ \bibinfo {year} {2003})\ pp.\ \bibinfo
  {pages} {200--209}\BibitemShut {NoStop}%
\bibitem [{\citenamefont {Chandrashekar}\ \emph {et~al.}(2008)\citenamefont
  {Chandrashekar}, \citenamefont {Srikanth},\ and\ \citenamefont
  {Laflamme}}]{chandrashekar2008optimizing}%
  \BibitemOpen
  \bibfield  {author} {\bibinfo {author} {\bibfnamefont {C.}~\bibnamefont
  {Chandrashekar}}, \bibinfo {author} {\bibfnamefont {R.}~\bibnamefont
  {Srikanth}}, \ and\ \bibinfo {author} {\bibfnamefont {R.}~\bibnamefont
  {Laflamme}},\ }\href@noop {} {\bibfield  {journal} {\bibinfo  {journal}
  {Physical Review A}\ }\textbf {\bibinfo {volume} {77}},\ \bibinfo {pages}
  {032326} (\bibinfo {year} {2008})}\BibitemShut {NoStop}%
\bibitem [{\citenamefont {Zhang}\ \emph {et~al.}(2016)\citenamefont {Zhang},
  \citenamefont {Goyal}, \citenamefont {Gao}, \citenamefont {Sanders},\ and\
  \citenamefont {Simon}}]{zhang2016creating}%
  \BibitemOpen
  \bibfield  {author} {\bibinfo {author} {\bibfnamefont {W.-W.}\ \bibnamefont
  {Zhang}}, \bibinfo {author} {\bibfnamefont {S.~K.}\ \bibnamefont {Goyal}},
  \bibinfo {author} {\bibfnamefont {F.}~\bibnamefont {Gao}}, \bibinfo {author}
  {\bibfnamefont {B.~C.}\ \bibnamefont {Sanders}}, \ and\ \bibinfo {author}
  {\bibfnamefont {C.}~\bibnamefont {Simon}},\ }\href@noop {} {\bibfield
  {journal} {\bibinfo  {journal} {New Journal of Physics}\ }\textbf {\bibinfo
  {volume} {18}},\ \bibinfo {pages} {093025} (\bibinfo {year}
  {2016})}\BibitemShut {NoStop}%
\bibitem [{\citenamefont {Di~Molfetta}\ \emph {et~al.}(2014)\citenamefont
  {Di~Molfetta}, \citenamefont {Brachet},\ and\ \citenamefont
  {Debbasch}}]{di2014quantum}%
  \BibitemOpen
  \bibfield  {author} {\bibinfo {author} {\bibfnamefont {G.}~\bibnamefont
  {Di~Molfetta}}, \bibinfo {author} {\bibfnamefont {M.}~\bibnamefont
  {Brachet}}, \ and\ \bibinfo {author} {\bibfnamefont {F.}~\bibnamefont
  {Debbasch}},\ }\href@noop {} {\bibfield  {journal} {\bibinfo  {journal}
  {Physica A: Statistical Mechanics and its Applications}\ }\textbf {\bibinfo
  {volume} {397}},\ \bibinfo {pages} {157} (\bibinfo {year}
  {2014})}\BibitemShut {NoStop}%
\bibitem [{\citenamefont {Genske}\ \emph {et~al.}(2013)\citenamefont {Genske},
  \citenamefont {Alt}, \citenamefont {Steffen}, \citenamefont {Werner},
  \citenamefont {Werner}, \citenamefont {Meschede},\ and\ \citenamefont
  {Alberti}}]{genske2013electric}%
  \BibitemOpen
  \bibfield  {author} {\bibinfo {author} {\bibfnamefont {M.}~\bibnamefont
  {Genske}}, \bibinfo {author} {\bibfnamefont {W.}~\bibnamefont {Alt}},
  \bibinfo {author} {\bibfnamefont {A.}~\bibnamefont {Steffen}}, \bibinfo
  {author} {\bibfnamefont {A.~H.}\ \bibnamefont {Werner}}, \bibinfo {author}
  {\bibfnamefont {R.~F.}\ \bibnamefont {Werner}}, \bibinfo {author}
  {\bibfnamefont {D.}~\bibnamefont {Meschede}}, \ and\ \bibinfo {author}
  {\bibfnamefont {A.}~\bibnamefont {Alberti}},\ }\href@noop {} {\bibfield
  {journal} {\bibinfo  {journal} {Physical review letters}\ }\textbf {\bibinfo
  {volume} {110}},\ \bibinfo {pages} {190601} (\bibinfo {year}
  {2013})}\BibitemShut {NoStop}%
\bibitem [{\citenamefont {Muraleedharan}\ \emph {et~al.}(2018)\citenamefont
  {Muraleedharan}, \citenamefont {Miyake},\ and\ \citenamefont
  {Deutsch}}]{muraleedharan2018quantum}%
  \BibitemOpen
  \bibfield  {author} {\bibinfo {author} {\bibfnamefont {G.}~\bibnamefont
  {Muraleedharan}}, \bibinfo {author} {\bibfnamefont {A.}~\bibnamefont
  {Miyake}}, \ and\ \bibinfo {author} {\bibfnamefont {I.~H.}\ \bibnamefont
  {Deutsch}},\ }\href@noop {} {\bibfield  {journal} {\bibinfo  {journal} {arXiv
  preprint arXiv:1805.01858}\ } (\bibinfo {year} {2018})}\BibitemShut {NoStop}%
\bibitem [{\citenamefont {Schwinger}(1960)}]{schwinger1960unitary}%
  \BibitemOpen
  \bibfield  {author} {\bibinfo {author} {\bibfnamefont {J.}~\bibnamefont
  {Schwinger}},\ }\href@noop {} {\bibfield  {journal} {\bibinfo  {journal}
  {Proceedings of the National Academy of Sciences}\ }\textbf {\bibinfo
  {volume} {46}},\ \bibinfo {pages} {570} (\bibinfo {year} {1960})}\BibitemShut
  {NoStop}%
\bibitem [{\citenamefont {Nayak}\ and\ \citenamefont
  {Vishwanath}(2000)}]{nayak2000quantum}%
  \BibitemOpen
  \bibfield  {author} {\bibinfo {author} {\bibfnamefont {A.}~\bibnamefont
  {Nayak}}\ and\ \bibinfo {author} {\bibfnamefont {A.}~\bibnamefont
  {Vishwanath}},\ }\href@noop {} {\bibfield  {journal} {\bibinfo  {journal}
  {arXiv preprint quant-ph/0010117}\ } (\bibinfo {year} {2000})}\BibitemShut
  {NoStop}%
\bibitem [{\citenamefont {Lin}\ and\ \citenamefont
  {Motrunich}(2018)}]{lin2018out}%
  \BibitemOpen
  \bibfield  {author} {\bibinfo {author} {\bibfnamefont {C.-J.}\ \bibnamefont
  {Lin}}\ and\ \bibinfo {author} {\bibfnamefont {O.~I.}\ \bibnamefont
  {Motrunich}},\ }\href@noop {} {\bibfield  {journal} {\bibinfo  {journal}
  {Physical Review B}\ }\textbf {\bibinfo {volume} {97}},\ \bibinfo {pages}
  {144304} (\bibinfo {year} {2018})}\BibitemShut {NoStop}%
\bibitem [{\citenamefont {Kukuljan}\ \emph {et~al.}(2017)\citenamefont
  {Kukuljan}, \citenamefont {Grozdanov},\ and\ \citenamefont
  {Prosen}}]{kukuljan2017weak}%
  \BibitemOpen
  \bibfield  {author} {\bibinfo {author} {\bibfnamefont {I.}~\bibnamefont
  {Kukuljan}}, \bibinfo {author} {\bibfnamefont {S.}~\bibnamefont {Grozdanov}},
  \ and\ \bibinfo {author} {\bibfnamefont {T.}~\bibnamefont {Prosen}},\
  }\href@noop {} {\bibfield  {journal} {\bibinfo  {journal} {Physical Review
  B}\ }\textbf {\bibinfo {volume} {96}},\ \bibinfo {pages} {060301} (\bibinfo
  {year} {2017})}\BibitemShut {NoStop}%
\bibitem [{\citenamefont {Riddell}\ and\ \citenamefont
  {Sorensen}(2018)}]{riddell2018out}%
  \BibitemOpen
  \bibfield  {author} {\bibinfo {author} {\bibfnamefont {J.}~\bibnamefont
  {Riddell}}\ and\ \bibinfo {author} {\bibfnamefont {E.~S.}\ \bibnamefont
  {Sorensen}},\ }\href@noop {} {\bibfield  {journal} {\bibinfo  {journal}
  {arXiv preprint arXiv:1810.00038}\ } (\bibinfo {year} {2018})}\BibitemShut
  {NoStop}%
\bibitem [{\citenamefont {Huang}\ \emph {et~al.}(2017)\citenamefont {Huang},
  \citenamefont {Zhang},\ and\ \citenamefont {Chen}}]{huang2017out}%
  \BibitemOpen
  \bibfield  {author} {\bibinfo {author} {\bibfnamefont {Y.}~\bibnamefont
  {Huang}}, \bibinfo {author} {\bibfnamefont {Y.-L.}\ \bibnamefont {Zhang}}, \
  and\ \bibinfo {author} {\bibfnamefont {X.}~\bibnamefont {Chen}},\ }\href@noop
  {} {\bibfield  {journal} {\bibinfo  {journal} {Annalen der Physik}\ }\textbf
  {\bibinfo {volume} {529}},\ \bibinfo {pages} {1600318} (\bibinfo {year}
  {2017})}\BibitemShut {NoStop}%
\bibitem [{\citenamefont {Sun}\ \emph {et~al.}(2018)\citenamefont {Sun},
  \citenamefont {Cai}, \citenamefont {Tang}, \citenamefont {Hu},\ and\
  \citenamefont {Fan}}]{sun2018out}%
  \BibitemOpen
  \bibfield  {author} {\bibinfo {author} {\bibfnamefont {Z.-H.}\ \bibnamefont
  {Sun}}, \bibinfo {author} {\bibfnamefont {J.-Q.}\ \bibnamefont {Cai}},
  \bibinfo {author} {\bibfnamefont {Q.-C.}\ \bibnamefont {Tang}}, \bibinfo
  {author} {\bibfnamefont {Y.}~\bibnamefont {Hu}}, \ and\ \bibinfo {author}
  {\bibfnamefont {H.}~\bibnamefont {Fan}},\ }\href@noop {} {\bibfield
  {journal} {\bibinfo  {journal} {arXiv preprint arXiv:1811.11191}\ } (\bibinfo
  {year} {2018})}\BibitemShut {NoStop}%
\bibitem [{\citenamefont {Omanakuttan}\ and\ \citenamefont
  {Lakshminarayan}(2018)}]{omanakuttan2018quantum}%
  \BibitemOpen
  \bibfield  {author} {\bibinfo {author} {\bibfnamefont {S.}~\bibnamefont
  {Omanakuttan}}\ and\ \bibinfo {author} {\bibfnamefont {A.}~\bibnamefont
  {Lakshminarayan}},\ }\href {http://stacks.iop.org/1751-8121/51/i=38/a=385306}
  {\bibfield  {journal} {\bibinfo  {journal} {Journal of Physics A:
  Mathematical and Theoretical}\ }\textbf {\bibinfo {volume} {51}},\ \bibinfo
  {pages} {385306} (\bibinfo {year} {2018})}\BibitemShut {NoStop}%
\bibitem [{\citenamefont {Lakshminarayan}(2003)}]{lakshminarayan2003random}%
  \BibitemOpen
  \bibfield  {author} {\bibinfo {author} {\bibfnamefont {A.}~\bibnamefont
  {Lakshminarayan}},\ }\href@noop {} {\bibfield  {journal} {\bibinfo  {journal}
  {arXiv preprint quant-ph/0305026}\ } (\bibinfo {year} {2003})}\BibitemShut
  {NoStop}%
\bibitem [{\citenamefont {Panahiyan}\ and\ \citenamefont
  {Fritzsche}(2018)}]{panahiyan2018one}%
  \BibitemOpen
  \bibfield  {author} {\bibinfo {author} {\bibfnamefont {S.}~\bibnamefont
  {Panahiyan}}\ and\ \bibinfo {author} {\bibfnamefont {S.}~\bibnamefont
  {Fritzsche}},\ }\href@noop {} {\bibfield  {journal} {\bibinfo  {journal}
  {arXiv preprint arXiv:1810.11020}\ } (\bibinfo {year} {2018})}\BibitemShut
  {NoStop}%
\bibitem [{\citenamefont {Lorz}\ \emph {et~al.}(2018)\citenamefont {Lorz},
  \citenamefont {Meyer-Scott}, \citenamefont {Nitsche}, \citenamefont
  {Potocek}, \citenamefont {G{\'a}bris}, \citenamefont {Barkhofen},
  \citenamefont {Jex},\ and\ \citenamefont {Silberhorn}}]{lorz2018photonic}%
  \BibitemOpen
  \bibfield  {author} {\bibinfo {author} {\bibfnamefont {L.}~\bibnamefont
  {Lorz}}, \bibinfo {author} {\bibfnamefont {E.}~\bibnamefont {Meyer-Scott}},
  \bibinfo {author} {\bibfnamefont {T.}~\bibnamefont {Nitsche}}, \bibinfo
  {author} {\bibfnamefont {V.}~\bibnamefont {Potocek}}, \bibinfo {author}
  {\bibfnamefont {A.}~\bibnamefont {G{\'a}bris}}, \bibinfo {author}
  {\bibfnamefont {S.}~\bibnamefont {Barkhofen}}, \bibinfo {author}
  {\bibfnamefont {I.}~\bibnamefont {Jex}}, \ and\ \bibinfo {author}
  {\bibfnamefont {C.}~\bibnamefont {Silberhorn}},\ }\href@noop {} {\bibfield
  {journal} {\bibinfo  {journal} {arXiv preprint arXiv:1809.00591}\ } (\bibinfo
  {year} {2018})}\BibitemShut {NoStop}%
\bibitem [{\citenamefont {Chen}\ \emph {et~al.}(2017)\citenamefont {Chen},
  \citenamefont {Zhou}, \citenamefont {Huse},\ and\ \citenamefont
  {Fradkin}}]{chen2017out}%
  \BibitemOpen
  \bibfield  {author} {\bibinfo {author} {\bibfnamefont {X.}~\bibnamefont
  {Chen}}, \bibinfo {author} {\bibfnamefont {T.}~\bibnamefont {Zhou}}, \bibinfo
  {author} {\bibfnamefont {D.~A.}\ \bibnamefont {Huse}}, \ and\ \bibinfo
  {author} {\bibfnamefont {E.}~\bibnamefont {Fradkin}},\ }\href@noop {}
  {\bibfield  {journal} {\bibinfo  {journal} {Annalen der Physik}\ }\textbf
  {\bibinfo {volume} {529}},\ \bibinfo {pages} {1600332} (\bibinfo {year}
  {2017})}\BibitemShut {NoStop}%
\end{thebibliography}%

\end{document}